# Measurement of the Mass Profile of Abell 1689

J. Anthony Tyson [1] and Philippe Fischer

AT&T Bell Laboratories, 600 Mountain Ave., Murray Hill, NJ 07974

## ABSTRACT

In this letter we present calibrated mass and light profiles of the rich cluster of galaxies Abell 1689 out to 1 $h^{-1}$ Mpc from the center. The high surface density of faint blue galaxies at high redshift, selected by their low surface brightness, are unique tools for mapping the projected mass distribution of foreground mass concentrations. The systematic gravitational lens distortions of $10^4$ of these background galaxies in 15' fields reveal detailed mass profiles for intervening clusters of galaxies, and are a direct measure of the growth of mass inhomogeneity. The mass is measured directly, avoiding uncertainties encountered in velocity or X-ray derived mass estimates.

Mass in the rich cluster Abell 1689 follows smoothed light, outside 100 $h^{-1}$ kpc, with a rest-frame V band mass-to-light ratio of $400 \pm 60$ $h^{-1}(M/L_V)_\odot$. Near the cluster center, mass appears to be more smoothly distributed than light. Out to a radius of 1 $h^{-1}$ Mpc the total mass follows a steeper than isothermal profile. Comparing with preliminary high resolution N-body clustering simulations for various cosmogonies on these scales, these data are incompatible with hot dark matter, a poor fit to most mixed dark matter models, and favor open or $\Lambda > 0$ cold dark matter. Substructure is seen in both the mass and the light, but detailed correspondence is erased on scales less than 100 $h^{-1}$ kpc.

*Subject headings:* dark matter - galaxies: clustering - gravitational lensing -

---

## 1. Introduction

Clusters of galaxies are among the largest overdensities in the universe, so that detailed knowledge of their total mass profile over a wide range of radii, particularly on scales less than 10 $h^{-1}$ Mpc where non-linear structures grow, can constrain the nature of dark matter. Different cosmogonic models predict observably different mass clumping on large and small scales. If the COBE normalization is used for 1000 Mpc scales, then 10 - 1000 kpc scale measurements of mass are a powerful discriminant of cosmogonic models. Inversion of gravitational arclets offers the only known direct way of measuring this projected mass overdensity as a function of radius. In contrast, indirect estimates of the underlying mass distribution via galaxy number density or velocity, or X-ray studies, suffer from systematic uncertainties.

The timescale for mass assembly in clusters is a large fraction of the Hubble time, so that observations of the cluster mass spectrum at moderate redshifts can probe the initial conditions and overdensity spectrum. How mass fluctuations in such extreme overdense regions join onto the field mass fluctuations bears directly on the process of structure formation (Press & Schechter 1974; Ostriker 1994). Comparison of the mass density profile with the rest-frame luminosity density profile is also of interest; in some scenarios a direct correspondence between mass and light might be expected, including that in which the dark matter is baryonic.

To explore this observationally we require calibrated projected mass and light observations over a wide range of cluster radii. Deep CCD observations of thousands of distorted background galaxies, and cluster light, over a large area provide the necessary data. Giant arcs at the Einstein radius (Soucail, et al. 1988; Lynds & Petrosian 1989) provide the calibration. Here we report the first calibrated measurement of the detailed mass profile of a cluster of galaxies. A more complete description of our techniques will be given in Fischer & Tyson (1995). We apply this tomographic inversion to 6000 blue arcs and arclets in a 15′ field centered on Abell 1689, a cluster which we studied earlier with a 4′ CCD (Tyson et al. 1990, hereafter referred to as TVW), and compare the surface mass density profile with preliminary high resolution N-body simulations.

## 2. Observations and Processing

We have constructed very deep $B_j$ and R images from many shift-and-stare 100 – 500s CCD exposures covering 16 – 250 arcmin$^2$ taken during three runs on the CTIO 4m from 1989 – 1993. The total exposure times were 3.5 hr in $B_j$ (62 frames) and 3.8 hr in R (58 frames). Since the seeing and signal-to-noise of the individual frames varied we used an optimal weighting algorithm to combine them (Fischer & Kochanski 1994). The RMS scatter in the sky backgrounds on 1″ scales in these reconstructed images is 28.7 $B_j$ mag arcsec$^{-2}$ and 28.2 R mag arcsec$^{-2}$. The stellar FWHM is 1.5″ in the combined $B_j$ image and 1.3″ FWHM in the R image. The reduced full-coverage field size of 15 × 15′ covers $1.8 \times 1.8 h^{-1}$ Mpc at the redshift of the cluster, z = 0.18 (Teague, et al 1990). A color image of the central 12′ is shown in Fig. 1.

The probability of a random source intersecting the critical line increases exponentially with limiting magnitude, increasing the likelihood of multiple strongly lensed images in very deep images. Two concentric long blue arcs at the Einstein radius subtending 31° and 37° can be clearly seen in Fig. 2, a color image of the inner 3′ region in which the bright cluster galaxies have been modeled and subtracted (ProFit: Fischer 1995). These two arcs have a common center of curvature which is consistent with the position of the central CD galaxy. Given reliable limits on their redshifts, the long arcs provide a self-calibration of the mass scale for the cluster. This calibration depends on the assumed mass profile interior to the Einstein ring but spans a small range for plausible profiles. Because of competing volume and faint-end luminosity function effects, galaxies at $B_j$ = 25 mag. cover a wide range in redshifts extending from z = 1–3; blue low surface brightness sources are at the highest redshifts. The trend towards redshift ≥ 1 at 25th $B_j$ magnitude is clear in a magnitude – $\log(z)$ plot (Koo & Kron 1992; Tyson 1995; Lilly 1995). Due to the relatively low redshift of Abell 1689, the mass calibration varies by only ±6% for arc redshifts ranging from z = 1 – 2. The *shape* of the derived mass profile is unaffected by this uncertainty.

## 3. Arclet Inversion

For each source galaxy image, we use $FOCAS$ to calculate the intensity-weighted second moment of the image orthogonal to and along the radius relative to the lens center. Let $a^2$ and $b^2$ be the principal axis transformed transverse and radial second moments of a background galaxy relative to point (x,y) on the sky, corrected for trailing. A dimensionless image distortion related to the projected mass overdensity $T(x,y,r) = (a^2 - b^2)/(a^2 + b^2)$, is defined for every background galaxy at position $\vec{r}$ relative to this point (x,y) in the image plane (TVW). In the weak lensing limit, and for a circularly symmetric lens, it can be shown that the average tangential alignment of sources at radius $r$ from the lens center, $\bar{T}(0,0,r)$, is a measure of the mass contrast inside radius r (Miralda-Escudé 1991):

$$\bar{T}(0,0,r) \;=\; 2\,[\bar{\Sigma}(<r) - \Sigma(r)]\,\Sigma_c^{-1}, \tag{1}$$

where $\Sigma_c$ is the critical density (Turner, Ostriker & Gott 1984). To construct a map of the gravitational lens a weighted distortion statistic $D(x,y) = <T(x,y,r)>_r$ is computed over a grid of candidate lens centers. At any point $\vec{u} = (x,y)$ in the image plane we can sum over the tangential alignments of all source images about that point, creating a smoothed scalar distortion statistic $D(\vec{u})$:

$$D(\vec{u}) = \int K(\vec{r})\, T(\vec{r} - \vec{u})\, d\vec{r}, \qquad (2)$$

where the apodization kernel $K(\vec{r})$ weights source images over a radius range. Eqn. 2 yields "alignment" images which closely reproduce complicated mass morphologies in multi-lens simulations (TVW, Tyson 1992). Kaiser and Squires (1993) showed that $K(r)$ must asymptotically approach the power law $r^{-2}$ at large $r$ in order that $D(r)$ approximate the mass. Smoothing is required; this may either be incorporated in the kernel K or done after the fact (as in TVW). In practice, less noisy maps are produced by replacing hard limits in the integral by a smooth top hat filter in the kernel. Near the edge of the image our kernel also rolls off, minimizing edge effects (TVW). Since light bending angles from different mass components add, the distortion $D(x,y)$ at any point is related to the mass density contrast if the mass can be represented as a sum of cylindrically symmetric distributions. Much less noisy maps of mass morphology may be made by going to a less singular kernel: While not a map of mass density, a distortion image D(x,y) with $K(r) \sim r^{-1}$, related to $M(<r, \theta)/r$, uniquely locates the lens mass and gives a much higher signal-to-noise ratio for its morphological shape on the sky than with $r^{-2}$ weighting.

By probing a grid of points on the sky for coherent arclet alignment equation 2 gives a useful map of the lens morphology, but an accurate and high S/N plot of the radial mass distribution of a given lens may be obtained by summing arclets about the known lens center (a variant of equation 1): it can be shown (Fischer & Tyson 1995) that the azimuthally averaged projected mass density interior to radius $r$ is given by

$$\bar{\Sigma}(r) = \frac{\Sigma_c C}{2(1 - r^2/r_{max}^2)} \int_r^{r_{max}} T(r)\, d\ln r + \bar{\Sigma}(r, r_{max}), \qquad (3)$$

where C is the seeing correction found in simulations of the data, and $\bar{\Sigma}(r, r_{max})$ is the average mass density in the annulus between $r$ and $r_{max}$.

We test the inversion software by simulating cluster lens distortions for various cluster mass profiles. Field and cluster galaxies are distributed in six redshift shells and have the same properties as the observed galaxies (number-magnitude counts, size-magnitude relationship, ellipticity distribution, spatial distribution etc.). After the distortion for each redshift shell is calculated, the image is convolved with the PSF, binned down to the CCD resolution, and photon noise is added. Lens inversion is performed on the resulting $FOCAS$ or $ProFit$ galaxy catalogs, and the result compared with the input mass profile of the lens. As shown in Fig. 3, the shape of the output profiles are in excellent agreement with the input profiles and exhibit no significant systematic biases. Furthermore, the profile shapes are relatively insensitive to the way in which the galaxies used in the inversion were culled, although the overall mass scale varied significantly and is also a function of seeing.

### 3.1. Cluster Mass Morphology

Based on the $B_j-R \sim 0-1$ colors of faint blue field galaxies, we select on magnitude, color and also for low surface brightness thus reducing contamination from cluster and other foreground galaxies. For the Abell 1689 field, data on 5950 low surface brightness blue arclets found in a 140 arcmin$^2$ area were passed to the apodized moment analysis. A small net shear over the field due to trailing was removed by subtracting the scaled average star moments (Valdes, et al. 1983). A rough map of the mass distribution in the field was made via equation 2 with $K(r) = r^{-1}\exp(-r^2/2r_{max})[1-\exp(-r^2/2r_{min}]$, with $r_{max}$ equal to half the field size and $r_{min} = 14''$. Fig. 4 shows this mass morphology in the field of the cluster Abell 1689; contours of the smooth distortion $D(x,y)$ are overlaid on a red image of the cluster. The lowest contour is not reliable. Note that the distribution is not symmetric, having an extension towards and beyond the sub-clump of red galaxies to the northeast. The centroid of the map is consistent with the center of the bright arcs.

### 3.2. Cluster Mass Radial Profile

Having found the center of the lens, equation 3 is used to derive an accurate radial mass profile out to 1 h$^{-1}$ Mpc. Mass calibration is obtained in two ways: (1) Einstein critical radius as given by two faint long arcs, and (2) simulations of the lens as discussed above. These agree within 10%, method (1) giving $1.8 \pm 0.1 \times 10^{14}$ h$^{-1}$ $M_\odot$ inside the critical radius of 51'' (100 h$^{-1}$ kpc) for arc redshifts between 1 and 2. This corresponds to an average mass density $\bar{\Sigma}(< r_E)$ interior to 100 h$^{-1}$ kpc of $1.17 \pm 0.07$ h g cm$^{-2}$ or $5.6 \pm 0.3 \times 10^3$ h $M_\odot$ pc$^{-2}$. Eqn. 3 is then used to calculate $\bar{\Sigma}(< r) - \bar{\Sigma}(r, r_{max})$ from ProFit catalogs of weak lensing arclet shapes. Normalization of this mass density contrast is obtained from the strong lensing result for $\bar{\Sigma}(< r_E)$ for some mass density profile; we use the Navarro et al. (1995) (NFW) profile (see below). A calibrated plot of this projected mass density contrast, along with the luminosity density contrast in the rest-frame V band, is shown in Fig. 5 as a function of radius out to 1 $h^{-1}$ Mpc. To 1 $h^{-1}$ Mpc the best fit power-law exponent for projected mass and light is $n = -1.4 \pm 0.2$, which is steeper than an isothermal profile ($n = -1.0$).

NFW observe that typical $\Omega=1$ CDM N-body simulations appear to roughly follow a one parameter family of mass profiles in 3-space:

$$\frac{\rho(R)}{\bar{\rho}} = \frac{1500 R_{200}^3}{R(5R + R_{200})^2}, \quad (4)$$

where $\bar{\rho}$ is the mean density of the universe at the epoch considered, R is the radius in 3-space, and $R_{200}$ is the radius within which the mean overdensity is 200. Projecting Eqn. 4 gives $R_{200} = 3.1^{+0.10}_{-0.15}$ $h^{-1}$ Mpc for Abell 1689, which leads to a profile which is flatter than isothermal in the range plotted in Fig. 5. The corresponding mass contained within the Abell radius of 1.5 $h^{-1}$ Mpc is $2.7 \times 10^{15} h$ M$_\odot$, about 6 times that estimated for Coma from kinematic data. Our mass density slope is insensitive to a wide range of galaxy culling criteria and, therefore, this disagreement is unlikely to be a result of systematic measurement error. A much better fit is obtained by introducing a separately adjustable amplitude in Eqn. 4, leading to $R_{200} = 0.7 \pm 0.3$ $h^{-1}$ Mpc. Soft core profiles like these often produce thin radial arcs inside the Einstein radius; we plan HST observations to study these and constrain the shape of the mass distribution interior to 100 $h^{-1}$ kpc.

## 4. Cluster Mass-to-Light Ratio

We measure the luminosity density of the cluster by removing bright stars and measuring the difference of averages inside and outside of a radius, $\bar{L}_R(<r) - \bar{L}_R(r, r_{max})$, identically to the mass density measurement. Calibration was performed during photometric weather using exposures in the $B_j RI$ calibration fields (Gullixson et al. 1995).

At $z = 0.18$, the R band exactly corresponds to the rest-frame V band. Model assumptions in the k-correction may thus be avoided; in addition to the $-2.5\log(1+z) = -0.18$ mag cosmological dimming term the color-redshift correction simplifies to a ratio of the flux of an A0 star at the $B_j$ and R wavelengths (+0.54 mag). The absolute magnitude in the rest-frame V band is then given by

$$M_V = m_R - (m-M)_{bol} + 0.36. \quad (5)$$

The total luminosity inside the Einstein radius R = 14.1 mag. For a luminosity distance of $D_L = 560h^{-1}$ Mpc, this corresponds to $M_V(< r_E) = -24.3$. The total rest frame luminosity inside the Einstein radius is $L_V(< r_E) = 4.44 \times 10^{11} h^{-2} L_{V\odot}$, including a 30% contribution from the diffuse flux. The mean surface brightness inside this radius is thus 14.3 $L_{V\odot}$ pc$^{-2}$. To compare directly with the mass profile, the cluster rest-frame light contrast $\bar{L}_V(< r) - \bar{L}_V(r, r_{max})$, calibrated as above, is plotted along with the mass density contrast in Fig. 5. On scales larger than $\approx 100h^{-1}$ Mpc, mass apparently follows smoothed light, with a rest-frame V band mass-to-light ratio of $400 \pm 60$ $h$ $(M/L_V)_\odot$.

## 5. Discussion

These calibrated data on the mass profile of a rich cluster of galaxies can be used to test theories of cosmology and structure formation. While some form of hot dark matter (HDM) may contribute to a uniform mass background, clusters of galaxies appear to have a relatively peaked mass distributions; the projected mass profile in Abell 1689 is marginally consistent with a power law of slope $-1.4\pm0.2$. Some N-body codes do not preserve adequate resolution, so we compare only with detailed simulations on scales 5 times larger than their resolution limit. The best NFW profile with $R_{200} = 3.1h^{-1}$ Mpc overestimates the mass density in the radial range of our observations. The two parameter NFW-like profile discussed in §3.2 is a better fit. Our data are inconsistent with the very extended cores found in high resolution HDM N-body simulations (Cen 1994, 1995). Mass profiles interior to 1 Mpc in HDM model simulations are very flat. Our data also appear to be inconsistent with mixed dark matter simulations with significant HDM component. Continuous accretion in standard CDM tends to increase the mass density at 1 Mpc, driving flatter interior profiles. In contrast, low $\Omega$ universes tend to produce compact massive clusters (West, et al. 1987); preliminary OCDM N-body simulations (Xu 1995) with 20 kpc resolution show steep projected mass density profiles of slope -1.4 to -1.7 in the region exterior to 100 kpc where details of N-body resolution are less important. Our observed mass profile is intermediate between SCDM and $\Omega = 0.35$ OCDM or $\Lambda$ CDM simulated profiles, but a better match to the latter two. The OCDM models have trouble producing clusters as massive as Abell 1689, and it is worth keeping in mind that Abell 1689 is not a typical cluster.

There is kinematic data available for 66 plausible cluster member galaxies (Teague et al 1990), yielding a velocity dispersion of $2355^{+238}_{-183}$ km s$^{-1}$, which, with the usual assumptions about isotropy, results in a mass estimate much higher than the gravitational lensing result. This is probably best explained as the result of contamination and/or subclustering for which there is evidence in both the galaxy distribution and the mass morphology (Fig. 1 and 4). Assuming isothermality and hydrostatic support, ROSAT X-ray data yield a projected mass density approximately two times lower than ours within the Einstein radius (Daines et al.

1995). Since we already have a high redshift for the arcs compared with the cluster redshift, it is not possible for us to lower our mass estimate substantially and this remains a significant disagreement.

Finally, we find that mass is correlated with light only for scales larger than about 100 $h^{-1}$ kpc. Mass clumped on smaller scales could not produce the long arcs, all segments of the same critical circle. The observed sub-structure in the light and mass, together with the larger scale correlation of azimuthally averaged mass and light, presents an interesting constraint on the formation of structure. The tail of mass to the NE may be the legacy of a previous merger; the mass deficit apparent at large radius in Fig. 5 and to the southwest in Fig. 4 may represent an accretion edge. Comparison of observed mass profiles *and morphologies* with future high resolution N-body simulations on these scales of non-linear growth, using a variety of codes, will be particularly useful in discriminating between the various cosmogonic models.

We gratefully acknowledge the help of Gary Bernstein, Pat Boeshaar, Raja Guhathakurta, Greg Kochanski, Jordi Miralda-Escudé, Rick Wenk and the expert staff at CTIO. We also thank Renyue Cen and Guohong Xu for sharing their N-body results with us before publication, and Craig Hogan, David Spergel and Ed Turner for helpful conversations. PF thanks NSERC and Bell Labs for postdoctoral fellowships.

---

Fig. 1.— [PLATE XXX] A deep color image of the Abell 1689 field, reconstructed from 7.3 hours of $B_j$ and R CCD exposures. Sky rms noise is 28.7 $B_j$ and 28.2 R mag arcsec$^{-2}$. North is up and east is left. This color image is $12'$ across.

Fig. 2.— [PLATE XXX] A deep color image of the inner 6% of the Abell 1689 field. All red cluster galaxies have been modeled and subtracted. Note the long blue arcs at the Einstein radius of $51''$ ($100h^{-1}$ kpc) and the diffuse red light.

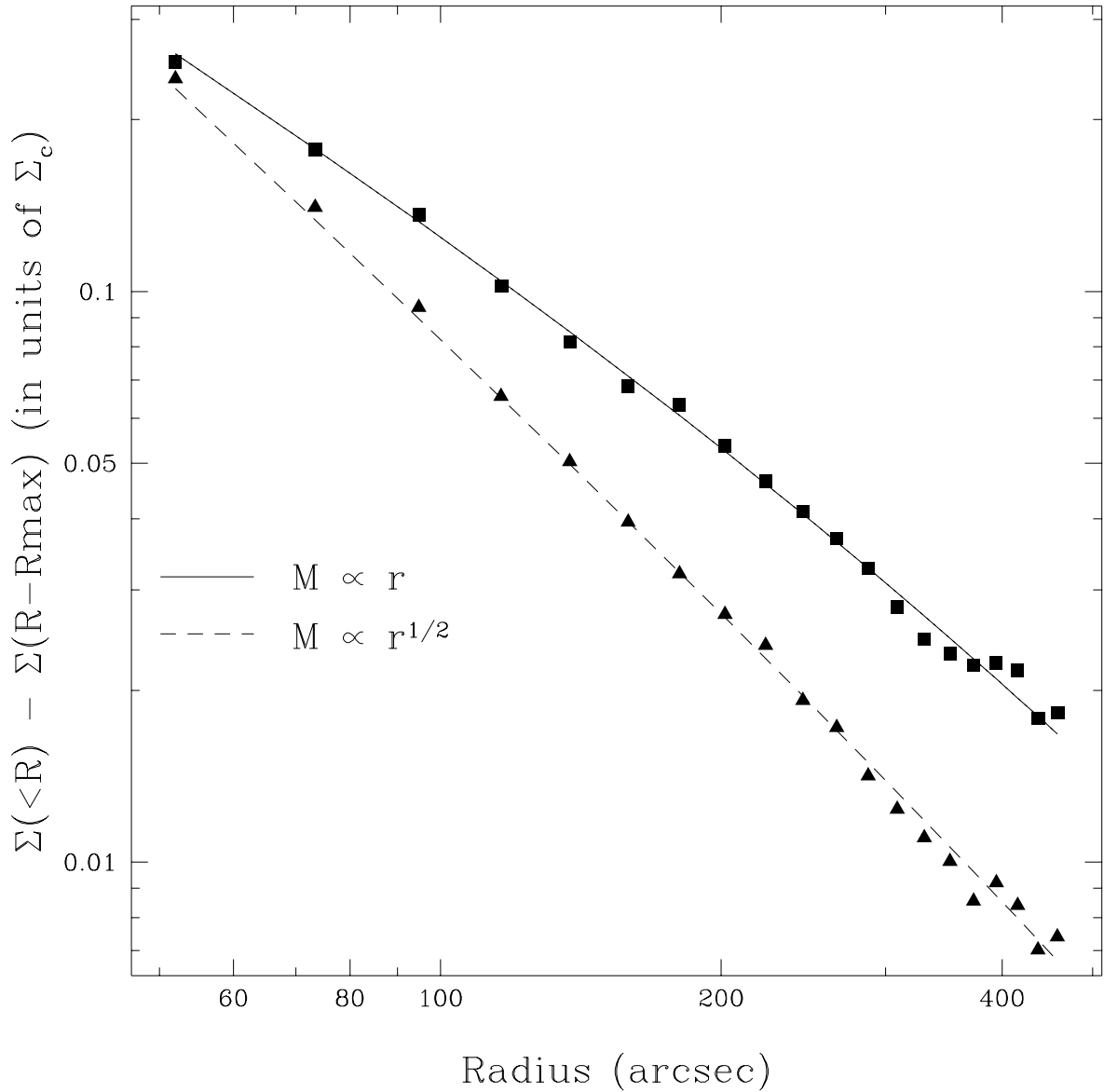

Fig. 3.— Input (solid lines) and output mass (points) density profiles for realistic simulations having characteristics similar the Abell 1689 data. These have had no seeing corrections applied. The input profiles have been scaled vertically to match the output profiles. The two output profiles represent averages of 39 and 26 simulations for the $M(r) = r^{-0.5}$ and $M(r) = r^{-1}$ models, respectively. No significant systematic biases are seen in the profile shape, however the mass scale is sensitive to seeing and culling of the galaxies.

Fig. 4.— A map of the smoothed gravitational lens distortion in the field of the $z = 0.18$ cluster Abell 1689 shown as contours plotted over the red image of the field. This extract from the full $1.8 \times 1.8 h^{-1}$ Mpc field is 1.50 $h^{-1}$ Mpc high; north is up and east is left. Mass and light are correlated on $> 100 h^{-1}$ kpc scales. The tail of mass to the NE may have resulted from a merger.

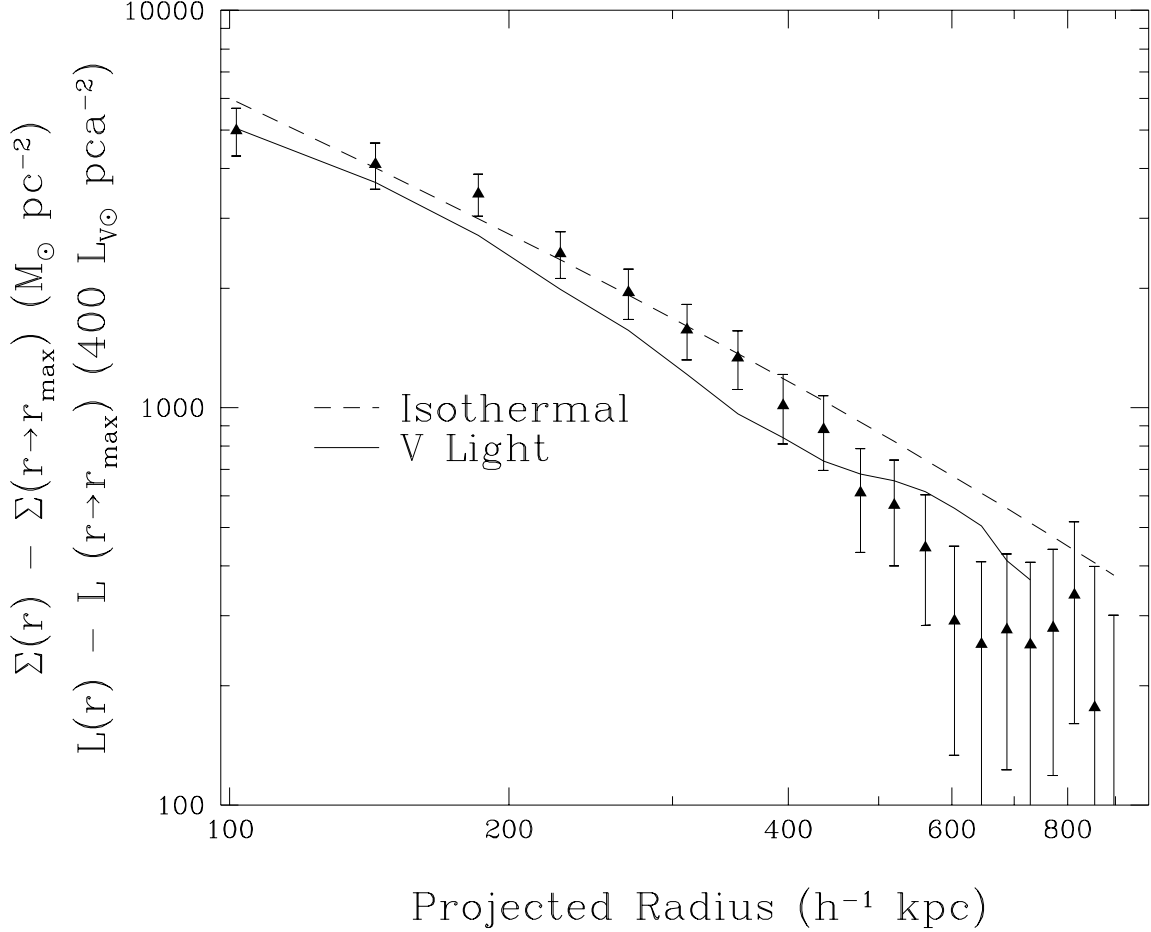

Fig. 5.— The projected azimuthally averaged profiles of the mass density contrast $\bar{\Sigma}(<r) - \bar{\Sigma}(r, r_{max})$ and the analogous luminosity density contrast in the rest-frame V band of Abell 1689. The error bars represent the scatter in the measurements and agree well with the values found in simulations. For comparison, an isothermal distribution (dashed line) is shown. HDM N-body simulations appear virtually flat on these scales; N-body open or closed CDM simulations are steeper than isothermal and are a better fit.